\documentclass{article}
\usepackage[dblblindworkshop, final]{neurips_2026}
\workshoptitle{Agentic AI for Biological Discovery}
\usepackage[utf8]{inputenc}
\usepackage[T1]{fontenc}
\usepackage{hyperref}
\usepackage{url}
\usepackage{booktabs}
\usepackage{amsfonts}
\usepackage{nicefrac}
\usepackage{microtype}
\usepackage{xcolor}
\usepackage{graphicx}
\usepackage{amsmath}

\title{Unlocking the Regulatory Genome by ARGUS: An Evidence-Constrained Agentic Framework for Interpreting Single Nucleotide Variants}

\author{%
  \textbf{Pratik Dutta}\textsuperscript{1} \quad
  \textbf{Matthew B. Obusan}\textsuperscript{2} \quad
  \textbf{Max Chao}\textsuperscript{1} \quad
  \textbf{Rekha Sathian}\textsuperscript{1} \\
  \textbf{Nimisha Papineni}\textsuperscript{1} \quad
  \textbf{Ramana V. Davuluri}\textsuperscript{1,2} \\[6pt]
  \textsuperscript{1}Stony Brook University, Stony Brook, NY, USA \\
  \textsuperscript{2}Renaissance School of Medicine at Stony Brook University, Stony Brook, NY, USA
}

\begin{document}

\maketitle

\begin{abstract}
Over 90\% of disease-associated variants from genome-wide association studies fall in noncoding regulatory regions, yet their functional interpretation remains a central open problem in genomic medicine. Large language models prompted to interpret such variants routinely hallucinate transcription factor (TF) binding changes, fabricate experimental support, and assign biological significance to statistically negligible signals. We present ARGUS (Agentic Regulatory Genomics for an Uncertainty-aware Scientist), which strictly separates deterministic biological computation from LLM-mediated reasoning. ARGUS wraps 458 DNABERT-based TF binding models in a hypothesis-directed investigation loop where a planner selects evidence sources based on current uncertainty, a verifier deterministically interprets each observation, and intermediate results change the investigation path. On variant rs6983267 at the 8q24 cancer risk locus, the same planner produces four divergent trajectories for four TFs. FOXA1 is rescued in 3 steps when real ADASTRA allele-specific binding data (15 experiments, FDR = 0.030) reveals a model false negative masked by saturation (both alleles predicted at near-maximal binding, leaving the model unable to resolve allelic differences). KLF6 traverses 8 steps across ADASTRA, JASPAR motif analysis, and ENCODE cCRE regulatory annotation before abstaining due to mixed indirect evidence. RAD21 abstains in 8 steps after ADASTRA returns a powered but nonsignificant allelic test (5 experiments, FDR = 0.65), and SP1, which shares FOXA1's saturated retained prediction, abstains because no direct experimental evidence exists at this locus. All trajectories are produced from real data against local ADASTRA, JASPAR, and ENCODE cCRE evidence sources, with no simulated observations. A comparison of fixed-priority and LLM-mediated planning shows that the LLM planner reaches identical verdicts with fewer tool calls by declining evidence that cannot resolve the claim under test. 

Code Availability: \url{https://github.com/duttaprat/ARGUS}.
\end{abstract}

\section{Introduction}

The interpretation of noncoding regulatory variants remains a central bottleneck in translating GWAS findings into biological mechanism~\cite{maurano2012, corpas2026agentic}.
A typical GWAS locus contains dozens of variants in linkage disequilibrium, most falling in noncoding regulatory regions where they may disrupt transcription factor binding sites, alter chromatin accessibility, or rewire enhancer-promoter contacts.
Computational tools that predict the functional impact of such variants, including sequence-based deep learning models~\cite{zhou2015deepsea}, have made substantial progress on the prediction problem. More recently, AlphaGenome Atlas~\cite{alphagenome2026} precomputes variant impact predictions for all 9 billion possible single nucleotide variants; however, prediction and interpretation remain distinct 
problems, and no existing resource distinguishes between ``a model predicts an effect'' and ``independent evidence supports this effect.'' However, \textit{interpreting} a predicted binding change is fundamentally different from \textit{predicting} one: interpretation requires integrating the prediction with tissue context, disease relevance, independent experimental evidence, and an honest assessment of uncertainty.

Recent work has explored agentic AI systems for biological research~\cite{li2026agentic, corpas2026agentic}, where LLM-based agents coordinate multi-step analyses by calling specialized tools, retrieving literature, and synthesizing results.
These systems face a distinctive failure mode: the same language fluency that makes LLMs effective narrators also makes them unreliable biological reasoners. Recent benchmarking has shown that general-purpose LLMs can produce inaccurate cancer-variant classifications and tend to overclassify variants
supported by weaker evidence~\cite{lin2025benchmarking}. In our own prototype, we observed a related failure when an LLM interpreted raw TF-binding model outputs. In an early prototype using Qwen 2.5 7B as both the reasoning and reporting engine, the system analyzed variant rs6983267 at the 8q24 cancer risk locus and reported that SP1 showed ``a gain in binding with log\_odds\_ratio of 3.02.''
Both binding probabilities (ref = 0.007, alt = 0.0009) sat far below the 0.5 active-binding threshold: there was no binding at either allele, and therefore nothing to gain or lose.
The LLM hallucinated a biologically significant event from a large ratio of two negligible numbers.

This experience crystallized a design principle we call \textbf{deterministic science, agentic reasoning}: all biological classification is performed by rule-based, auditable code; the LLM may select investigation actions or write the final report, but it never classifies evidence or assigns verdicts, constrained by pre-classified facts it cannot override. ARGUS extends this principle with a hypothesis-directed investigation loop in which a planner autonomously selects evidence sources, a verifier deterministically interprets each observation, and intermediate results change the investigation path. The defining criterion for agency in this system is that \textit{an observation changes what ARGUS does next}.

\section{Architecture}

ARGUS consists of two stages: a deterministic prediction stage (shared with the existing DeepVRegulome pipeline~\cite{deepvregulome} and orchestrated with LangGraph~\cite{langgraph}) and an agentic investigation stage that subjects each prediction to evidence-based scrutiny.

\begin{figure}[t]
\centering
\includegraphics[width=\textwidth]{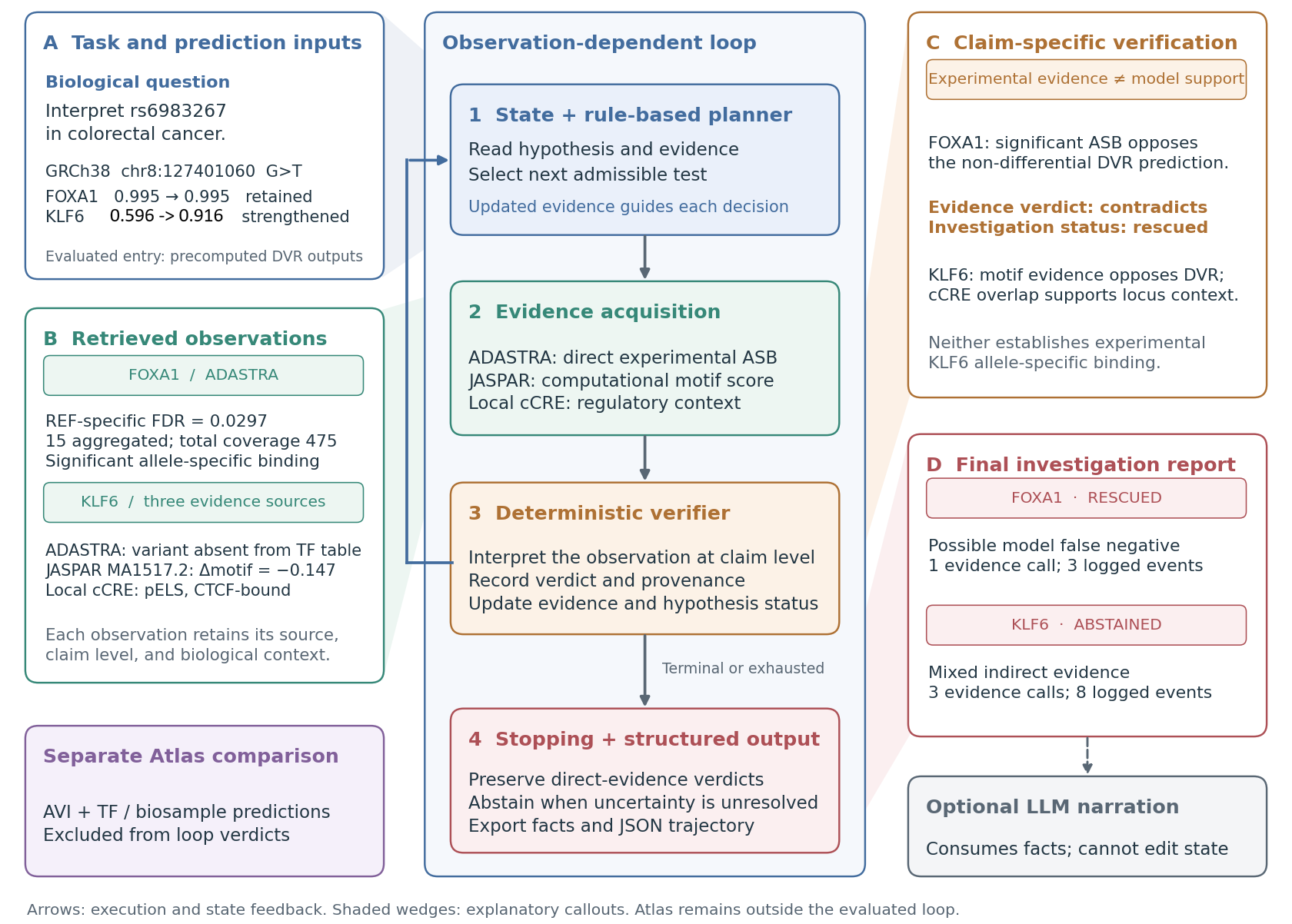}
\caption{ARGUS architecture. Left: task inputs with precomputed DVR predictions. Center: the observation-dependent investigation loop (state/planner, evidence acquisition, deterministic verifier, stopping rule). Right: claim-specific verification showing how FOXA1 (rescued, 3 steps) and KLF6 (abstained, 8 steps) reach different verdicts from the same planner responding to different intermediate observations. AlphaGenome Atlas remains outside the evaluated loop.}
\label{fig:arch}
\end{figure}

\subsection{Prediction Stage}

User queries are routed to one of four modes via deterministic regex matching (no LLM involvement): variant mode, gene mode, region mode, and TF mode.
For each variant, ARGUS selects TFs from its library of 458 DNABERT-based binding models using a curated knowledge base cross-referencing tissue-specific expression (GTEx), ChIP-seq peak overlap (193M+ peaks across 700 TFs from ENCODE and ChIP-Atlas), and disease-relevant regulatory annotations.
Each selected TF model receives a 510-bp sequence centered on the variant, with ref and alt alleles substituted, and outputs binding probabilities $p_{\text{ref}}$ and $p_{\text{alt}}$.

A deterministic classifier assigns one of four labels based on a 0.5 active-binding threshold: \textbf{LOF/GOF} (one allele $\geq 0.5$, the other $< 0.5$; the variant crosses the binding threshold), \textbf{binding strengthened/weakened} (both alleles $\geq 0.5$ but quantitatively shifted), \textbf{binding retained} (both alleles $\geq 0.5$ with negligible change), or \textbf{no confident binding} (both alleles $< 0.5$).
The classifier also flags \textbf{saturation} when both probabilities exceed 0.95, indicating the model has reached its dynamic range ceiling.
This classification structurally prevents the SP1 hallucination: a large log-odds ratio between two sub-threshold probabilities cannot produce a binding-change label.

\subsection{Investigation Stage: The Agentic Loop}

Each classifier output initializes a \texttt{HypothesisState} object that tracks: the TF-variant identity, the DVR prediction, the current epistemic status (\texttt{active}, \texttt{supported}, \texttt{contradicted}, \texttt{rescued}, \texttt{abstained}, or \texttt{partially\_supported}), all evidence collected so far, and a complete decision trajectory.

The investigation loop is a stateful cycle, not a fixed directed acyclic graph.
Writing the current state as $S_t$ and the enabled tool set as $\mathcal{T}$, each step proceeds as:
\begin{equation}
a_t = \pi(S_t, \mathcal{T}), \qquad o_t = T_{a_t}(S_t), \qquad S_{t+1} = V(S_t, o_t),
\end{equation}
where $\pi$ is the rule-based planner policy, $T_{a_t}$ is the selected evidence tool, and $V$ is the deterministic verifier.
The policy may also select a stopping action, terminating the loop.

Five modules implement this cycle with strictly separated responsibilities:

\textbf{State} (\texttt{state.py}) stores scientific facts: DVR predictions, evidence records, hypothesis status, and trajectory events.
It makes no decisions about what to test next or what an observation means.

\textbf{Planner} (\texttt{planner.py}) selects the next evidence source based on current uncertainty.
It implements a priority-ordered policy: ADASTRA (experimental allele-specific binding) is queried first for all hypotheses, including null predictions.
If ADASTRA does not resolve the hypothesis, the planner selects JASPAR (allele-specific motif scoring).
If JASPAR does not resolve it, ENCODE cCRE (regulatory element context) is queried.
If no admissible evidence source remains, the planner returns \texttt{ABSTAIN}. An alternative LLM-mediated planner is also implemented: Claude Sonnet 4.6 reads the current hypothesis state, remaining budget, and available tool descriptions via structured tool use, then proposes the next action with an explicit rationale. Both planners share the same verifier and stopping rule; only the action-selection policy differs.
The planner prioritizes hypotheses by risk: saturated predictions (highest), followed by LOF/GOF, quantitative shifts, retained binding, and null predictions.

\textbf{Tools} execute real evidence queries.
Three tools are currently wired:
\begin{itemize}
    \item \textbf{ADASTRA} (Mabel v6.1.1): local lookup against bulk TSV files containing allele-specific binding events from thousands of ChIP-seq experiments~\cite{abramov2021adastra}. Returns FDR-corrected p-values, experiment counts, read coverage, preferred allele, and motif concordance. Absence from a queried TF table is recorded as unavailable evidence, not as proof of absent coverage elsewhere.
    \item \textbf{JASPAR} (2024 release): allele-specific motif scoring using the JASPAR CORE vertebrate PSSM library~\cite{jaspar2024}. Scores ref and alt sequence windows at the variant position; directional differences exceeding an absolute threshold of 0.05 in relative score are compared with the DVR prediction direction. This is a heuristic scoring threshold, not a significance test. Motif evidence is computational and cannot independently establish or refute experimental TF binding. Directional comparison is applied only when DVR predicts a direction; for non-directional predictions (retained or no binding), motif evidence is recorded as unavailable for the claim.
    \item \textbf{ENCODE cCRE}: local tabix lookup against indexed GRCh38 ENCODE SCREEN candidate cis-regulatory elements~\cite{encode2020}. Input positions are 1-based; the lookup uses the 0-based, half-open interval $[p{-}1, p)$. Reports whether the variant overlaps an annotated regulatory element (promoter-like, enhancer-like, CTCF-bound, or DNase-only). A failed query is recorded as unavailable evidence; it is not converted into ``no overlap.''
\end{itemize}

Of the 458 TF models in the ARGUS library, 445 have matching ADASTRA tables after mapping gene symbols to UniProt entry names, defining the current scope of direct experimental falsification.

\textbf{Verifier} (\texttt{verifier.py}) deterministically interprets each tool's raw output and updates the hypothesis state.
For ADASTRA, the verifier implements a 10-branch decision tree handling: no data (unavailable; stay active), significant ASB with saturated model (contradicts DVR null; rescued), significant ASB contradicting a null prediction (contradicts; contradicted), significant ASB with direction matching the prediction (supports; supported), significant ASB with direction mismatch (contradicts; contradicted), non-significant with adequate power (contradicts for differential predictions; neutral for non-differential predictions, because nonsignificance does not establish absence of an allelic effect), and non-significant with inadequate power (underpowered; stay active).
For JASPAR, evidence is tagged as computational and indirect (\texttt{experimental=False}, \texttt{direct\_to\_claim=False}); it can weaken or support a hypothesis but never independently confirm or falsify TF binding.
ENCODE cCRE provides regulatory context only.

\textbf{Stopping rule} (\texttt{stopping.py}) determines the final status when the planner returns \texttt{ABSTAIN}.
Consistent indirect support produces \texttt{partially\_supported}.
Mixed indirect evidence (some sources support, others contradict) produces \texttt{abstained}.
Underpowered direct evidence produces \texttt{abstained}.
No resolving evidence produces \texttt{abstained}.
The stopping rule never creates \texttt{supported}, \texttt{contradicted}, or \texttt{rescued}; those stronger verdicts belong exclusively to the verifier acting on direct experimental evidence.

\subsection{Evidence-Constrained Reporter}

In the default configuration, the reporter is the only component that uses an LLM; the optional LLM planner (Section 3.5) selects actions but cannot modify evidence, verdicts, or state transitions. It receives a structured facts block from the completed investigation containing: variant annotations, classified binding results with arm labels, evidence records with verdict and provenance tags, and explicit \texttt{CAN\_CLAIM\_EXPERIMENTAL} flags. The LLM generates narrative text from these pre-classified facts; it cannot override any verdict, reclassify any binding event, or claim experimental confirmation where \texttt{CAN\_CLAIM\_EXPERIMENTAL} is \texttt{False}. Constraining state transitions does not guarantee factuality of unrestricted downstream prose; full narrative evaluation is deferred to future work.

\section{Results}

\subsection{Divergent Trajectories at rs6983267}

We evaluated ARGUS on variant rs6983267 (chr8:127401060 G$>$T, GRCh38) at the 8q24 cancer risk locus.
Four TFs were investigated, producing four divergent trajectories from the same planner operating on real evidence.
All observations are from actual local database queries; no results are simulated.

\begin{table}[h]
\centering
\caption{ARGUS investigation trajectories for four TFs at rs6983267. Each trajectory is produced by the same planner responding to different intermediate observations.}
\label{tab:trajectories}
\small
\begin{tabular}{@{}llcllp{4.5cm}@{}}
\toprule
\textbf{TF} & \textbf{DVR Arm} & \textbf{Steps} & \textbf{Tools Used} & \textbf{Final Status} & \textbf{Key Observation} \\
\midrule
FOXA1 & retained [sat.] & 3 & ADASTRA & rescued & Significant ASB (FDR = 0.030, 15 exp., 475 reads); contradicts DVR's non-differential prediction \\
KLF6 & strengthened & 8 & ADASTRA, JASPAR, cCRE & abstained & ADASTRA unavailable; JASPAR motif opposes DVR ($\Delta = {-}0.147$); cCRE supports (pELS); mixed evidence \\
RAD21 & retained [sat.] & 8 & ADASTRA, JASPAR, cCRE & abstained & Powered but nonsignificant ASB (FDR = 0.65, 5 exp., 104 reads); no RAD21 motif; cCRE context only \\
SP1 & retained [sat.] & 8 & ADASTRA, JASPAR, cCRE & abstained & No ADASTRA data at locus; motif direction not adjudicated for non-directional arm; cCRE context only \\
\bottomrule
\end{tabular}
\end{table}

\textbf{FOXA1 (rescued, 3 steps).}
The DNABERT model predicts $p_{\text{ref}} = 0.995$ and $p_{\text{alt}} = 0.995$, both above the 0.95 saturation threshold.
The classifier assigns \texttt{binding\_retained} with a saturation warning.
The planner recognizes saturation and queries ADASTRA.
Real ADASTRA data returns significant allele-specific binding for FOXA1 at rs6983267 (preferred allele: ref, FDR = 0.030, 15 ChIP-seq experiments, 475 reads).
The verifier records this as contradicting DVR's non-differential prediction (\texttt{relation\_to\_dvr = discordant}) and assigns \texttt{rescued}: a possible model false negative where independent experimental data indicates a real allelic event the model cannot distinguish.
The planner stops because the hypothesis has reached a terminal state.

\textbf{KLF6 (abstained, 8 steps).}
DVR predicts strengthened KLF6 binding ($p_{\text{ref}} = 0.596$, $p_{\text{alt}} = 0.915$).
The planner queries ADASTRA: KLF6 has an ADASTRA file but rs6983267 is not among the tested SNPs.
The verifier records \texttt{unavailable} and the hypothesis remains \texttt{active}.
The planner, observing that ADASTRA did not resolve the hypothesis, selects JASPAR as the next orthogonal test.
JASPAR returns a discordant result: motif MA1517.2 shows $\text{ref\_rel} = 0.725$ and $\text{alt\_rel} = 0.577$ ($\Delta = -0.147$), meaning the alt allele \textit{decreases} the KLF6 motif score, opposite to DVR's prediction of strengthened binding.
The verifier tags this as \texttt{contradicts} (computational, not experimental) and the hypothesis remains \texttt{active}.
The planner, observing that JASPAR contradicts the prediction, queries ENCODE cCRE for regulatory context.
The local cCRE lookup finds the variant overlaps a proximal enhancer-like signature (pELS) with CTCF binding (accessions EH38D5988281, EH38E3858447).
The verifier records \texttt{supports} (contextual, not experimental).
With no remaining evidence sources, the planner returns \texttt{ABSTAIN}.
The stopping rule observes mixed indirect evidence (JASPAR contradicts, cCRE supports) and assigns \texttt{abstained}.
\texttt{CAN\_CLAIM\_EXPERIMENTAL} is \texttt{False}: ARGUS refuses to promote this prediction to a finding.

\textbf{RAD21 (abstained, 8 steps).}
DVR predicts retained binding at saturation ($p_{\text{ref}} = 0.999$, $p_{\text{alt}} = 0.9995$).
The planner queries ADASTRA, which returns a nonsignificant allele-specific test that passes the coverage filter (5 experiments, 104 reads, FDR = 0.65).
Because nonsignificance does not establish the absence of an allelic effect, the verifier records this result as neutral and the hypothesis remains active.
JASPAR contains no matching RAD21 motif, and ENCODE cCRE places the variant in a proximal enhancer-like, CTCF-bound element, which provides context only.
With only contextual support, the stopping rule assigns \texttt{abstained}.

\textbf{SP1 (abstained, 8 steps).}
DVR predicts retained binding at saturation ($p_{\text{ref}} = p_{\text{alt}} = 0.9995$). ADASTRA contains an SP1 table, but rs6983267 is not among its tested SNPs. JASPAR finds the SP1 motif MA0079.5 with a lower ALT score ($\Delta = -0.120$), but the verifier does not adjudicate motif direction against a non-directional prediction, so this observation is recorded as unavailable for the claim. The cCRE overlap provides context only, and the stopping rule assigns \texttt{abstained}. SP1 and FOXA1 enter the loop with the same saturated, retained prediction; FOXA1 is rescued because direct experimental evidence exists, whereas SP1 abstains because it does not.

\subsection{Observation-Dependent Branching}

The key result is that the same planner, given the same variant and the same tool set, produces structurally different trajectories because intermediate observations differ.
FOXA1 resolves in one tool call (ADASTRA is definitive).
KLF6 requires all three tools because each intermediate result leaves the hypothesis unresolved: ADASTRA has no data, JASPAR contradicts, and cCRE provides partial support. RAD21 and SP1 follow the same tool path as KLF6 but stop for different reasons: a powered nonsignificant experimental test, and the absence of direct evidence, respectively. The planner's decision at each step depends on the verdict of the previous step, not on a predetermined sequence. The planner's decision at each step depends on the verdict of the previous step, not on a predetermined sequence.

This conditional branching distinguishes ARGUS from a fixed pipeline.
A pipeline runs all tools regardless of intermediate results; ARGUS stops when evidence is sufficient and continues when it is not.
The 3-step FOXA1 trajectory and the 8-step KLF6 trajectory are produced by the same code path; only the observations differ.

\subsection{Frontier LLM Baseline}

When prompted to interpret rs6983267, frontier LLMs (GPT-4o, Claude 4.6 Sonnet, Gemini 2.5 Pro) produce confident analyses that conflate well-known 8q24/MYC biology with variant-specific TF binding claims, cite experimental literature without verifying variant-level resolution, and make no distinction between ``a TF is expressed in this tissue'' and ``this TF shows allele-specific binding at this exact position.''
None spontaneously identifies the limits of its knowledge, distinguishes prediction from evidence, or abstains from unverifiable claims.
This is an informal observation, not a controlled benchmark; systematic LLM comparison across a broader variant set is deferred to future work.

\subsection{AlphaGenome Atlas Cross-Reference}

To assess concordance with an independent computational model, we queried the AlphaGenome Atlas (released September 8, 2026)~\cite{alphagenome2026} for rs6983267 using SDK version 0.9.0.
The AlphaGenome Variant Impact (AVI) score is 0.570, indicating high predicted regulatory impact.
Among 1,617 returned CHIP\_TF tracks, TF-specific predictions show the largest absolute change for FOXA1 (max $|\Delta| = 0.463$ across 4 tracks including HepG2 and MCF-7), consistent with the ADASTRA finding, and a negative KLF6 change ($\Delta = -0.240$, 1 genetically modified HepG2 track), directionally concordant with JASPAR's motif discordance.
RAD21 has 16 tracks (max $|\Delta| = 0.107$) and SP1 has 8 tracks (max $|\Delta| = 0.206$), both including HCT116 (a colorectal cancer cell line).
These are computational predictions with substantial biosample dependence; they are excluded from the planner and verifier decisions because the AVI score is variant-level (identical for all TFs) and would not produce observation-dependent branching.
Integration of TF-specific directional predictions as a fourth evidence source is planned.

\subsection{Fixed vs.\ LLM-Mediated Planning}

To assess whether the fixed evidence priority (ADASTRA, JASPAR, cCRE) captures the decisions a reasoning agent would make, we ran all hypotheses through both the fixed planner and an LLM-mediated planner (Claude Sonnet 4.6, budget of 3 tool calls per hypothesis). The LLM planner receives the current hypothesis state, the remaining tools with one-line capability descriptions, and the remaining budget; it returns a structured action proposal with an explicit rationale. All biological classification remains deterministic regardless of which planner is used.

\begin{table}[h]
\centering
\caption{Fixed vs.\ LLM planner comparison across five hypotheses on identical frozen evidence snapshots. Tool columns show the sequence each planner selected, with tool calls used out of a budget of 3 in parentheses. LLM results were identical across three independent replicates.}
\label{tab:planners}
\small
\begin{tabular}{@{}llllll@{}}
\toprule
\textbf{TF} & \textbf{Variant} & \textbf{Fixed tools} & \textbf{LLM tools} & \textbf{Fixed status} & \textbf{LLM status} \\
\midrule
FOXA1 & rs6983267 & A (1/3) & A (1/3) & rescued & rescued \\
RAD21 & rs6983267 & A, J, C (3/3) & A, J (2/3) & abstained & abstained \\
KLF6  & rs6983267 & A, J, C (3/3) & A, J (2/3) & abstained & abstained \\
SP1   & rs6983267 & A, J, C (3/3) & A, J (2/3) & abstained & abstained \\
BACH1 & rs2981578 & A, J, C (3/3) & A, J (2/3) & abstained & abstained \\
\bottomrule
\end{tabular}

\vspace{4pt}
{\footnotesize A = ADASTRA, J = JASPAR, C = ENCODE cCRE.}
\end{table}

Both planners reached identical verdicts in all five cases, and the LLM planner selected the same tool sequence in each of three independent replicates.
The planners diverged only in budget use.
For the four abstaining hypotheses, the LLM planner stopped after ADASTRA and JASPAR, declining to query ENCODE cCRE, whereas the fixed planner queried all three sources.
Because cCRE overlap establishes regulatory context but cannot resolve TF-specific binding or its direction, it could not change any of these verdicts; the LLM planner reached the same conclusions with 9 tool calls instead of 13.
This comparison shows that LLM-mediated selection can avoid evidence that is uninformative for the claim under test, while biological adjudication remains unchanged because both planners share the same deterministic verifier and stopping rule. Each LLM planner call consumed approximately 1,200 input tokens and 250 output tokens (approximately \$0.01 per hypothesis at current API pricing), with a full audit trail (input state, model response, rationale) saved in the trajectory JSON. With five hypotheses and one alternative policy, this comparison illustrates budget behavior; it does not establish a general planning advantage.

\section{Discussion}

\subsection{Architecture Enforces What Prompting Cannot}

The central lesson of ARGUS is that the boundary between ``what the LLM may do'' and ``what it may not do'' must be enforced architecturally, not by prompting.
Prompt-level instructions to ``only report significant findings'' are insufficient: LLMs can and do override such instructions when the input signal fits a pattern they associate with significance.
By performing all biological classification in deterministic code and passing only pre-labeled facts to the LLM, ARGUS eliminates the category of hallucinations where the model misinterprets numerical outputs as biologically meaningful.

\subsection{Prediction vs.\ Interpretation}

ARGUS does not compete with variant effect prediction models such as DeepVRegulome, AlphaGenome, or Enformer.
It addresses the downstream problem: given a prediction from any such model, should we believe it?
This distinction is important because variant effect prediction has received substantial investment (AlphaGenome Atlas precomputes predictions for 9 billion SNVs), but the interpretation problem, which requires integrating predictions with independent evidence, assessing uncertainty, and knowing when to abstain, has received comparatively little attention in the agentic AI literature.
ARGUS is model-agnostic: its planner, verifier, and stopping rules apply regardless of the upstream prediction source.

\subsection{Falsification as a Design Pattern for Agentic Science}

This workshop asks how we should build systems that go beyond static prediction toward closed-loop biological discovery~\cite{xin2025agentic}. ARGUS offers a concrete design pattern: treat every prediction as a testable hypothesis, select evidence sources based on unresolved uncertainty, interpret each observation through deterministic rules, and make the system's epistemic state explicit at every step. The categorical evidence states (supported, contradicted, rescued, abstained) communicate not just what the system predicts, but what it has tested and what remains unknown.

The distinction between experimental and computational evidence is enforced at the data structure level.
ADASTRA evidence carries \texttt{experimental=True, direct\_to\_claim=True}; JASPAR motif evidence carries \texttt{experimental=False, direct\_to\_claim=False}. The reporter's \texttt{CAN\_CLAIM\_EXPERIMENTAL} flag is \texttt{True} only when direct experimental evidence supports the hypothesis. This prevents a common failure mode in agentic systems where computational concordance is presented as experimental confirmation.

\subsection{Limitations}

The evaluation comprises five TF hypotheses at two loci. It demonstrates the system's behavior and trajectory divergence but does not constitute a predictive accuracy benchmark or a controlled measurement of hallucination reduction. The initial DVR outputs are precomputed; the loop has not been evaluated as an end-to-end natural-language-to-inference system.

The current planner implements a fixed evidence priority (ADASTRA, JASPAR, cCRE) rather than a learned or value-of-information-based policy. Adaptive planning, formal stopping criteria based on expected information gain, and a regulatory hypothesis graph that tracks cross-TF dependencies are identified as future extensions.

The falsification battery is limited to three evidence sources.
Expansion to eQTL concordance (GTEx), chromatin accessibility (ATAC-seq), literature evidence (PubMed), and independent model predictions (AlphaGenome Atlas CHIP\_TF) would broaden the epistemic reach.

The DNABERT models have known blind spots, including saturation at high binding probabilities and limited sensitivity to variants outside core motif positions.
The saturation rescue path addresses the most severe of these, but systematic characterization of model failure modes remains ongoing.
Biological interpretation is further limited by tissue matching between the model training context and evidence sources, assay aggregation in ADASTRA, and reference/alternate allele harmonization.

ENCODE cCRE data was queried from locally indexed BED files because the SCREEN API was unreachable from the compute environment; results are from the same underlying data but lack biosample-specific activity annotations available through the API.

\section{Conclusion}

ARGUS demonstrates that the hallucination problem in agentic genomics is not an inevitable cost of using LLMs for biological interpretation.
By separating deterministic science from agentic reasoning, embedding falsification as a first-class component, and making every verdict traceable to a specific evidence observation, the system produces conclusions that are not merely plausible but actively tested.
The four divergent trajectories at rs6983267, produced from real evidence by the same planner responding to different intermediate observations, demonstrate that observation-dependent investigation is achievable within a principled, auditable framework.

\begin{ack}
\textbf{Funding.} This work was financially supported by the National Library of Medicine of the National Institutes of Health (R01LM01372201 to R.D.). We also thank Anthropic for providing API credits through the Anthropic AI for Science program, which supported the LLM planner and reporter experiments.

\textbf{Competing interests.} The authors declare no competing interests.
\end{ack}

\bibliographystyle{unsrtnat}

\end{document}